\documentstyle[aps]{revtex}
\def \be {\begin{equation}}
\def \ee {\end{equation}}

\begin{document}

\title{The evolution of primordial black hole masses 
in the radiation-dominated era}

\author{Paulo Sergio Custodio and J.E. Horvath}
\footnote{Instituto de Astronomia, Geof\'\i sica  e Ci\^encias Atmosf\'ericas, 
Universidade de S\~ao Paulo, Av. M. St\'efano 4200, Agua Funda, 
04301-904 S\~ao Paulo, Brazil}

\clearpage
\maketitle

\begin{abstract}

We discuss in this work the behaviour of primordial black holes
(PBHs) in the radiative era.  Taking into account the Hawking
evaporation and the absorption of energy we revisit the complete
differential equation for the evolution of the mass of a PBH. We
show that the mass can grow in this cosmological phase in a very
slow fashion (even when considering the very high temperature of the
radiation) if at all, and give a strong upper limit to the maximum
accretion of mass. We evaluate relativistic effects due to the
peculiar motion relative to the CMBR and show that the existence
of relativistic black holes with very high mass absorption is
highly unlikely. Finally we demonstrate that thermodynamical
equilibrium between black holes and the cosmic radiation can not
exist for finite times, and therefore initially non-evaporating 
PBHs must jump to the evaporating regime. This analysis supports the 
several efforts performed to look for signatures of evaporating holes.

\end{abstract}

\section{Introduction}

The search for PBHs is certainly one of the most interesting areas
of high-energy astrophysics. While quite strong limits have been
published \cite{LG} on the fractional mass in PBHs, it is possible that
a fraction of short GRBs may correspond to evaporating holes
having masses at formation of the order the Hawking mass
$M\sim{10}^{15}g$ that would be evaporating today \cite{CG}. The celebrated 
Hawking radiation would be driving black holes to their end, although 
several important details regarding this final phase are still being 
analyzed. However, Sivaram \cite{Siv}
has argued that these objects would not be observed at all, since
that these PBHs would have absorbed a substantial amount of mass when the
Universe was very young and hot, and therefore would not be
evaporating today. This cautionary remark is important, since then short
GRBs must be produced by other sources. Here we reanalyze the
problem, and show that substantial mass gain did not occur at all
(Section 2). The kinematic analysis of Section 3 indicates that a
possible boost to enhance the mass absorption term is severely
suppressed. The last Section is devoted to a demonstration of the
impossibility of keeping thermodynamic equilibrium between the
black holes and the radiation, thus justifying the previous
discussion and the expected evolution of the mass. A slightly 
more general argument along these lines is given in the Appendix.

\section{ Evaporation and absorption of mass}

Hawking \cite{SH}, found in a seminal paper 
that black holes would loss their mass
through a slow but irreversible quantum process, where virtual
particles from the vacuum are converted in real ones created by
the large gravitational energy. This quantum process gives rise to
mass loss where the black hole had its temperature increased
continuously. The mathematical analysis of this problem, showed
that this emitted spectrum is thermal, with an associated
temperature,

\be
T_{bh}={\frac{\hbar {c}^{3}}{8\pi{G}k_{B}M}} \sim {\frac{{10}^{-7}K}{(M/M_{\odot})}} \,\, .
\ee

Using the Stefan-Boltzmann law and the formula above, the evolution of the mass of the PBH is given by

\be
{\frac{dM}{dt}}=-{\frac{A(M)}{M^{2}}} ; 
\ee

where $A(M)$ counts the degrees of freedom of the emitted particles. Numerically, $A(M)$ is given by

\be
A(M)\sim{10}^{24}{g}^{3}{s}^{-1} \eta(M) \,\,; 
\ee

with $\eta(M)=7.8 \times{D_{1}(M)}+3\times{D_{1/2}(M)}$, $D_{1/2}$ and $D_{1}$ are the
multiplicity of quantum numbers (spin, charge and color) of the emitted particles. For the
standard model with three fermionic generations, we have $D_{1}=27$ and
$D_{1/2}=90$ at $T\sim{100}GeV$ and $A(M)\leq{7.8\times{10}^{26}g^{3}{s}^{-1}}$
(ignoring contributions from Higgs scalars and gravitons, see \cite{KT} for details).

We know that the cosmological environment was very hot and dense
in the radiation era, therefore, we expect some classical
absorption of the energy-matter from the surroundings of a given
PBH.  We adopt a Friedmann-Robertson-Walker model, with $R(t)$ the
scale factor describing the expansion (or contraction) of the
universe and $K=-1,0,+1$ are the possible values of the curvature
parameter, describing, respectively open models ($K=-1$ or $K=0$)
and closed models ($K=1$). The dynamics of the FRW model is 
governed by the well known equations

\be
{\biggl(\frac{\dot{R}}{R}\biggr)}^{2}={\frac{8\pi{G}}{3}}{\varrho}-{\frac{K}{R^{2}}} \,\,; 
\ee

\be
{\frac{\ddot{R}}{R}} = -\frac{4 \pi G}{3}(\varrho+3P) \,\, ; 
\ee

without cosmological constant (which is not relevant for our discussion). In the radiation-dominated era the universe was close to $K = 0$
and the scale factor  and the radiation temperature evolve as $R(t)=R_{i} \times {t}^{1/2}$ and
$T_{rad}(t)\propto{R(t)}^{-1}$ respectively.
Numerically, the temperature is given by $T_{rad}\sim{1MeV}{(t/1s)}^{-1/2}$ and then the
radiation density is given by

\be
\varrho_{rad}(t)\sim{8.4\times{10}^{4}}{(t/1s)}^{-2}g{cm}^{-3} \,\, . 
\ee

We define as usual the cosmological horizon by the integral
$R_{hor}(t)=R(t)\lim_{t_{i}\rightarrow{0}}\int_{t_{i}}^{t}{\frac{cdt^{\prime}}{R(t^{\prime})}}$.
The cosmological horizon exists when this limit converges for finite $t$. Taking the limit
$t_{i}\rightarrow{0}$ we have for the solution $R(t)\propto{t}^{1/2}$

\be
R_{hor}(t)= 2 c t \,\, . 
\ee

 The
mass contained in the horizon is given simply by the product
$M_{hor}(t)\sim{\frac{4\pi}{3}}{R_{hor}(t)}^{3}\varrho(t)$. Using
eqs.(6) and (7) we obtain

\be
M_{hor}(t)\sim {7.6 \times{10}^{37}}(t/1s)\, g \,\, .
\ee

This relation is very important to the problem, since we expect that PBHs would be formed
with masses comparable to the horizon mass \cite{Carr}. There is also a dependence with the adiabatic index of the background matter $\Gamma={\frac{P}{\varrho}}$, an important
feature for the collapse of perturbations associated to PBHs and generally overlooked. Since PBHs with
initial masses larger than the horizon mass are excluded by causality, for a generic PBH
with initial mass $M$ was formed in the early universe its formation time must satisfy

\be
t_{f}(M) \, \geq \, 3 \times {10}^{-38}{(M/g)} \, s \,\, . 
\ee

Therefore, from $t_{f}(M)$, we expect that the PBH mass will be below the horizon mass at subsequent times $t>t_{f}$.

In order to address the complete problem of the evolution of PBHs masses, we shall include a
classical absorption term. This term must be included to the r.h.s. of eq.(2) and is of form

\be
{\biggl({\frac{dM}{dt}}\biggr)}_{abs}=\sigma_{g}(M)F_{rad}(T) \,\, ;
\ee

where $\sigma_{g}(M)={\frac{27\pi}{4}}{r_{g}}^{2}$ is the gravitational cross section for the capture
of relativistic particles \cite{ZN}. Since  $F_{rad}(T)=c\varrho_{rad}(T)$ is the radiative flux as seen in
the rest frame of the black hole, the complete differential equation for the mass is given by

\be
{\frac{dM}{dt}}=-{\frac{A(M)}{M^{2}}}+{\frac{27\pi{G}^{2}}{c^{3}}}\varrho_{rad}(T){M}^{2} \,\, ;
\ee

where we have assumed that $M > M_{Planck}$ in order to avoid complicated corrections due to quantum gravity
effects and neglected the
back-reaction (see \cite{Norma} and references therein for a quantum mechanical
evaluation).

The inclusion of the second term is a necessary  ingredient for the thermodynamical description,
since PBHs formed obeying the causality constraints  with masses $\geq M_{Haw}$ before $t\sim{1s}$
will experience an inward heat flux from the  hotter surroundings.
Consequently, these  cold PBHs will gain mass-energy at the expense of the radiation. When the physical
temperature of these PBHs equals to the radiation temperature, these objects will be in instantaneous
equilibrium and $\dot{M}=0$ at this instant. We can derive the mass for equilibrium to happen
at any epoch solving $\dot{M}=0$ \cite{NOS}

\be
M_{c}(t)\sim{\frac{{10}^{26}g}{(T/T_{0})}} \,\, ;
\ee

where $T_{0}$ is the present cosmic temperature. Any PBH with mass greater than this value at $t$ must
be accreting ($\dot{M}( M > M_{c})>0$) and not evaporating, since the Hawking evaporation is negligible in this case.

To be quantitative we may consider the ratio of the absorption to evaporation terms to obtain

\be
\xi(M)={\frac{{\dot{M}}_{abs}}{{\dot{M}}_{evap}}}={\biggl({\frac{M}{M_{c}}}\biggr)}^{4} \,\, ;
\ee

which shows that the Hawking term quickly becomes negligible for masses above $M_{c}$.

A first look at eq.(13) could convince us to accept the
conclusions of Sivaram \cite{Siv}, namely that these supercritical  ($M > M_{c}$)
PBHs would be growing very fast in the early universe. To show
that this would be misleading we shall solve the eq.(11) for
initially supercritical PBHs, (that is, ignoring the first term) in
the thermal radiation bath evolving as stated in eq.(6). Then,
the differential equation to solve is

\be
{\frac{dM}{dt}}={\frac{27\pi{G}^{2}}{c^{3}}}\varrho_{rad}(t)M^{2} \,\, .
\ee

The integration of this equation is immediate and we  obtain

\be
{\frac{1}{M(t)}}={\frac{1}{M_{i}}}+{1.2 \times{10}^{-39}}{\biggl({{\frac{1}{t}}-{\frac{1}{t_{i}}}}\biggr)}{g}^{-1} \,\, ; 
\ee

where $M_{i}$ is the initial mass (at formation) and $t_{i}$ is the formation time (measured 
in seconds) for this object.
This solution is valid if the initial mass satisfies $M_{c}(t_{i})< M_{i}\leq M_{hor}(t_{i})$, as stated above.

We may invert the analytical solution eq.(15) above, casting it in the  form

\be
M(t>>t_{i})={\frac{M_{i}}{[1-\epsilon(M_{i},t_{i})]}} \,\, ; 
\ee

with $\epsilon(M_{i},t_{i})= 1.2 \times{10}^{-39}{\frac{(M_{i}/g)}{(t_{i}/1s)}}$.

Since $\epsilon(M_{i},t_{i})<<1$ for PBHs satisfying
$M_{i}<<M_{hor}(t_{i})$; and even in the most favourable
circumstances  $\epsilon(M_{i},t_{i}) \sim M_{hor}(t_{i})$, we
can write eq.(16) expanding it in a Taylor series and find that
the growth of these objects is $M \, \sim \,
{M_{i}}[1+\epsilon(M_{i},t_{i})]$. Note that the mass gain is
larger for larger masses, and also increases for early formation
times for a given mass. In fractional terms

\be
{\frac{\delta{M}}{M}}={\frac{M-M_{i}}{M_{i}}}=\epsilon(M_{i},t_{i}) \,\, . 
\ee

Note that the fractional mass gain saturates with time for $t>>t_{i}$, 
as expected from the time
dependence of the radiation density $\varrho_{rad}(t)\propto{t}^{-2}$. Even at $t = t_{i}$ the universe
was not dense enough to drive a very strong growing. We can show it in the following way. Suppose that
the hot environment  can indeed support a relativistic PBH mass gain of the gravitational radius
$\dot{r_{g}}\sim{c}$.  Let us show that the actual mass accretion was many orders of magnitude below
this value, i.e. $\dot{M}<<{\frac{c^{3}}{2G}}\sim 2 \times {10}^{38}g{s}^{-1}$.
For this purpose, assume that initially, we had
$\dot{M}={\frac{27\pi{G}^{2}}{c^{3}}}\varrho_{rad}M^{2}\sim{\frac{c^{3}}{2G}}$. Solving it, we find
the minimal density of the radiation
capable to drive the relativistic growing for that mass $M$

\be
\varrho_{min} \sim 5.7 \times{10}^{51}g{cm}^{-3}{(M/M_{Haw})}^{-2} \,\, . 
\ee

If  we compare $\varrho_{min}$ with eq.(6) we obtain that these
PBH would have to be formed {\it before} $t_{rel}(M)\sim{3.8
\times{10}^{-24}}{(M/M_{Haw})}s$. However, according to eq.(9)
causal PBH formation is allowed {\it after} 
$t_{f} > 3\times{10}^{-23}{(M/M_{Haw})}s$. Thus, most of the time
the actual growing has to be much smaller than the maximal
theoretical value $\dot{M}_{max}={\frac{c^{3}}{2G}}$, which needs
very extreme conditions.

A brief relativistic gain may nevertheless be allowed, since $t_{f}$ and $t_{rel}$ are very close. This means that
a PBH with horizon mass, formed exactly at $t_{f}$ will gain some mass, but the maximal fractional gain
is limited to be $\leq 0.04$. Therefore, if we consider the maximal gain, a PBH formed with 
$M_{i} \simeq {10}^{15} g$
would finish its evolution with final mass (at the end of the radiation era) greater than the Hawking mass, but
with a small gain. However, that is enough to delay its evaporation, since the evaporation timescale
is given by

\be
t_{evap}(M)\sim{t_{0}}{(M/M_{Haw})}^{3} \,\, ; 
\ee

therefore the mass $M\sim{M_{i}+0.04 M_{i}}$ then we obtain $t_{evap}(M)\sim{(1.04)}^{3}t_{0}$. 

The situation is quite different for PBHs with initial mass of order $M_{i}$ smaller than $M_{Haw}$. In this case,
even the maximal gain can not drive these PBHs above the Hawking mass at all. Therefore, these objects will show very
small modifications of their behaviour.
Note that the maximum gain is easily obtained for $M_{i} \sim {M_{hor}}$. Then substituting we have for that absolute maximal gain

\be
\epsilon_{max}(M_{i}\rightarrow{M_{hor}}) \sim 1.17 \times{10}^{-39}M_{i}
\biggl[ \frac{1}{3\times{10}^{-38}M_{i}} \biggr] \leq {0.04} \,\, .
\ee

In order to understand the full behaviour of the PBHs masses it is useful to consider the evolution of the critical mass.
The latter may be expressed in the form

\be
M_{c}(t) \sim  10 {M_{Haw}}{(t/1s)}^{1/2} \,\, . 
\ee

If we compare the behaviour of a PBH with initial mass above the critical mass (but obviously below the horizon mass)
and the growth of the critical mass, we observe that this object will cross the critical mass some time after its formation. While this PBH stays above the critical mass, it will not evaporate, but its final mass (that is, before the crossing time) will be given by eq.(19). The {\it crossing time} $t_{cross}$ is defined by

\be
M(t)=M_{i}[1+\epsilon(M_{i},t_{i})]\sim 10 M_{Haw}{(t_{cross}/1s)}^{1/2} \,\, ; 
\ee

with solution

\be
t_{cross}(M_{i})\sim{10}^{-2}s{(M_{i}/M_{Haw})}^{2}
{[1+\epsilon(M_{i},t_{i})]}^{2} \,\, . 
\ee

This explains why these objects did not gain much mass: the time interval in which these objects stay above
the critical mass is very small if compared to cosmological times, or to the timescale for evaporation
$t_{evap}\propto{t_{0}{M}^{3}}$.

Any supercritical PBH will be driven by the second term in eq.(14) and any subcritical
PBH will be driven by the first term, the Hawking evaporation.
In the section 3 we will show that the objects crossing the critical mass curve can not stay there, and must drift to the evaporation region of Fig.1 in which the evaporation quickly dominates.


\bigskip

\section{Mass gain induced by high proper motions}

\bigskip
Since we expect that at least for large Lorentz factors the absorption and evaporation processes must be modified,
we want to consider corrections due to the proper motion of the PBH relative to the
background radiation. This effect is described by the following expression

\be
T_{rad}(\theta,v)={\frac{T_{rad}(\theta,v=0)}{\gamma(1-\vec{n}.\vec{\beta})}} \,\, ; 
\ee

where we will denote $T_{rad}(\theta,v=0)$  simply as $T_{rad}$, the temperature of the background radiation
in the rest frame of the PBH when this PBH has $v=0$. Then, $T_{rad}(\theta,v)$ for arbitrary $v$, denotes the
angular distribution of the background radiation seen from the rest frame of this PBH when it moves with peculiar
velocity given by $v$. The angle $\theta$ is defined in the scalar product $\vec{n}.\vec{\beta}=v{\beta}cos(\theta)$.
$\vec{n}$ is the vector defining the sight of view for an arbitrary observer located at the Schwarzschild radius,
$\gamma={\frac{1}{\sqrt{1-{\beta}^{2}}}}$ and $\beta=(v/c)$.

The radiation density has an angular distribution  as seen from a
stationary observer located at the black hole horizon. The sign in
the denominator of eq.(24) is such that we have a blueshift in
the direction of the motion and a redshift in the opposite. The
absorbed radiation is an angular average and it is given by \cite{NOS}

\be
<\varrho_{rad}(T_{rad})>={\frac{1}{4\pi}}
\int{d\Omega{\varrho_{rad}(T_{rad}(\theta,\phi))}} \,\, . 
\ee

This integral is performed to give

\be
\biggl({4{\gamma}^{2}-\frac{1}{\gamma}}\biggr)\varrho_{rad}(T_{rad}(t)) \,\, ; 
\ee

where the time dependence in $T_{rad}(t)$ arises from the cosmological expansion, i.e. $T_{rad}\propto{R}^{-1}$.

Inserting this result in the eq.(11), we arrive at the
differential equation for the mass of a black hole moving with
velocity $v$, as seen by an asymptotic observer. Evidently $\varrho_{rad}$ in eq.(26) must be the same quantity given by eq.(6). Now, we seek a $\gamma$ capable to drive the rate of
mass accretion to the maximal rate
$\dot{M}_{\gamma}\sim \frac{c^{3}}{2G}$. Scaling the black hole
mass to Hawking mass units

\be
{\biggl({4{\gamma}^{2}-\frac{1}{\gamma}}\biggr)}
=\frac{7\times{10}^{53}}{{(M/M_{Haw})}^{2}}{(\varrho_{rad}(t)/g{cm}^{-3})}^{-1} \,\, . 
\ee

This is a second degree equation for $\gamma$
with solution

\be
\gamma(t)={\frac{\alpha}{8{\mu}^{2}\varrho_{rad}(t)}}
\biggl[{1+\sqrt{1+{\frac{16}{{\alpha}^{2}}}{M}^{4}{\varrho_{rad}(t)}^{2}}}\biggr] \,\, ; 
\ee

where $\alpha=7\times{10}^{53}$, $\mu \equiv {M/M_{Haw}}$ and
$\varrho_{rad}(t)$ is expressed in $g{cm}^{-3}$. Substituting
$\varrho_{rad}(t)=\varrho(t_{f}){(t/t_{f})}^{-2}$,  we can rewrite
eq.(28) as

\be
\gamma(t)={\frac{\alpha{(t/t_{f})}^{2}}{8{\mu}^{2}\varrho_{rad}(t_{f})}}
\biggl[{1+\sqrt{1+{\frac{16}{{\alpha}^{2}}}{\mu}^{4}{\varrho_{rad}(t_{f})}^{2}
{(t/t_{f})}^{-4}}}\biggr] \,\, ; 
\ee

since $\varrho_{rad}(t)\sim {8.4 \times{10}^{4}}{(t/1s)}^{-2}$, and
our solution holds only for $t > t_{f}(M)$ we obtain the
following inequality

\be
\varrho_{rad}(t\geq{t_{f}}) \leq 4.4 \times \frac{{10}^{51}}{\mu^{2}} \,\, . 
\ee

Therefore, the prefactor in eq.(28) has to be greater than $\sim 18 $,
for any $t\geq{t_{f}}$. We may rewrite the required Lorentz factor
as

\be
\gamma(t)>  {18}
\biggl[{1+\sqrt{1+{{\frac{16}{{\alpha}^{2}}}\mu^{4}{\varrho_{rad}(t)}^{2}}}}\biggr] \,\, .
\ee

Applying the same arguments to second term, we can finally write

\be
\gamma(t)= {18}{(t/t_{f})}^{2}
\biggl[1+\sqrt{1+6\times{10}^{-4}{(t/t_{f})}^{-4}}\biggr] \,\, . 
\ee

For asymptotic times $t>t_{f}$, we have $\gamma(t) \rightarrow 36 {(t/t_{f})}^{2}$.

Therefore, we conclude that even at the beginning of its life ($t
\sim {t_{f}}$) and when the PBH was formed with $M \sim
{M_{hor}}$,  a high value for the proper velocity $v$ would be
needed to drive the PBH into the fast growing mode (differing from
$c$ in one part per thousand). The additional terms in the square
brackets are positive and enhance the value of the minimum
$\gamma$ in eq.(31). The physical reason is very simple: from the
initial time formation, the radiation density decreases with time,
and consequently we will need to impart very high values for the
proper velocity to drive an object above the critical mass. This
particular situation was certainly unlikely in the early universe,
and we do not expect to have PBHs with substantial accretion of
mass. It is worth to remark that the ingredient that allowed us to
perform these numerical evaluations and conclusions is the concept
of critical mass (see also \cite{NOS2} for a related discussion).

\section{Impossibility of thermodynamical equilibrium between PBHs and the background radiation}

In this section we address the problem whether black holes and the surrounding radiation can be
in thermodynamical equilibrium. This would imply that the PBH mass to be constant. Page \cite{Dany}
analyzed this problem in a {\it gedanken} experiment with one black hole contained in a finite box
plus thermal radiation at a constant volume. It is easy to understand that this black hole within this
box will settle down in stationary equilibrium, with its mass remaining constant in time. In this case,
the total energy of the radiation $+$ black hole system is

\be
E=A T^{4}V_{box}+M_{bh} \,\, . 
\ee

Since that this system is closed, we have for variations in mass and temperature

\be
\delta{M}=-4 A V_{box}T^{3}\delta{T} \,\, . 
\ee

If the black hole loses mass due to Hawking effect $\delta{M}<0$,
then the radiation temperature must raise $\delta{T_{rad}}>0$ and
drive the black hole into equilibrium. If the black hole absorbs
too much mass the radiation temperature diminishes substantially
and additional accretion will not be allowed. We see immediatly
that the situation is more complicated if the box has variable
volume. It is immediate to conclude that microscopic variations in
the critical mass must obey

\be
\delta{M_{c}}\propto {(T_{rad})}^{-5}{\frac{\delta{M}}{V_{box}}} \,\, . 
\ee

For very high temperatures, the variations in the critical mass are very small, as expected.

We shall now analyze the more complex and realistic situation, consisting in black holes and radiation
within a cosmological model. For this, we emphasize that the critical mass is a cosmological quantity,
defined by the radiation temperature. Numerically, we have

\be
M_{c}(t) \sim \frac{10^{26}g}{(T_{rad}(t)/T_{0})} \,\, . 
\ee

Therefore, since the temperature decreases with cosmological time,
the critical mass grows monotonically. However, we note that the
critical mass allows an {\it instantaneous} equilibrium, since
that $\dot{M}(M = M_{c})=0$. As the universe expands a black hole
initially in equilibrium with the background radiation must
deviate from it because of the different temporal behaviours.

To prove this statement, let us evaluate the variation of the area of the black hole.
We know that $R_{g}={\frac{2GM}{c^{2}}}$, then the horizon area is given by $A=4\pi{R_{g}}^{2}$, and therefore we have

\be
dA={\frac{32{\pi}}{c^{4}}}G^{2}M dM  \,\, .
\ee

The sign of the mass variation is related to the energy that crossed the event horizon as $dE=c^{2}dM$.
Then an energy flux through the event horizon induces a variation in the area given by

\be
{\frac{dE}{dA}}={\frac{c^{6}}{32\pi{G}^{2}M}} \,\, . 
\ee

Then, the time derivative of the energy flux per unit area will be given by

\be
\biggl({\frac{d^{2}E}{dtdA}}\biggr)=-{\frac{c^{6}}{32\pi{G}^{2}M^{2}}}{\frac{dM}{dt}} \,\, . 
\ee

Substituting eq.(11) into eq.(39), we obtain

\be
\biggl({\frac{d^{2}E}{dtdA}}\biggr)={\frac{c^{6}}{32\pi{G}^{2}}}
\biggl[{\frac{A}{M^{4}}}-{\frac{27\pi{G}^{2}}{c^{3}}}\varrho_{rad}(t)\biggr] \,\, .
\ee

We will denote eq.(40) as $\sigma(M,t)$ for simplicity. We see that the energy is zero when $M = M_{c}$,
then $\sigma(M = M_{c},t)= 0$ for any $t$. However, by evaluating the time derivative at $M = M_{c}$ and we
will prove that the first derivative at the critical mass is not zero, indicating an inevitable deviation
from the initial equilibrium.

Note that $\sigma(M,t)$ is positive-defined for hot black holes (subcritical black holes) and negative
in the opposite case. Now, the time derivative of $\sigma(M,t)$ at the critical mass is given by

\be
\dot{\sigma}(M=M_{c},t)=-{\frac{27}{32}}c^{3}\dot{\varrho}_{rad}(t)>0 \,\, ;
\ee

and it is positive-defined since that the radiation density
decreases with time. If we integrate eq.(40) with the initial
condition $M = M_{c}(t_{i})$ until $t > t_{i}$ and given that the
radiation diluted during this interval due to expansion, we obtain
the result

\be
\int_{\sigma(M,t)}^{\sigma(M_{c},t_{i})}{d\sigma}=-{\frac{c^{6}}{32\pi{G}^{2}}}
\int_{t_{i}}^{t}dt
\biggl[{{\frac{4A}{M^{5}}}{\frac{dM}{dt}}+{\frac{27\pi{G}^{2}}{c^{3}}}\dot{\varrho}_{rad}(t)}\biggr] \,\, ; 
\ee

and the right side can be rearranged in the form
$-{\frac{c^{6}A}{8\pi{G^{2}}}}\int_{M_{c}}^{M}{\frac{dM}{M^{5}}}-{\frac{27}{32}}c^{3}
\int_{\varrho_{rad}(t_{i})}^{\varrho_{rad}(t)}d\varrho_{rad}$. Then we arrive at

\be
\sigma(M,t)=\sigma(M=M_{c},t_{i})+{\frac{Ac^{6}}{32\pi{G}^{2}}}
\biggl[{{\frac{1}{{M}^{4}}}-{\frac{1}{{M_{c}}^{4}}}}\biggr]-{\frac{27}{32}}c^{3}\Delta{\varrho_{rad}} \,\, . 
\ee

The first term $\sigma(M = M_{c},t_{i})= 0$ by definition, and the last term is negative as we already know.
If we consider that $M = M_{c}+\delta{M}$, the term in brackets can be evaluated to give

\be
\biggl[{\frac{1}{{M}^{4}}}-{\frac{1}{{M_{c}}^{4}}}\biggr]={\frac{1}{{M_{c}}^{4}}}
\biggl[-1+{\frac{1}{1+F(M_{c},\delta{M_{c})}}}\biggr] \,\, ;  
\ee

where $F(M_{c},\delta{M_{c}})={({\frac{\delta{M}}{M_{c}}})}^{4}
+6{({\frac{\delta{M}}{M_{c}}})}^{2}+4{({\frac{\delta{M}}{M_{c}}})}^{3}+4{({\frac{\delta{M}}{M_{c}}})}$.
Therefore, the term in brackets is always negative, for arbitrary variations in $\delta{M}$, after the PBH
had crossed the critical mass at $t_{i}$. Note that $F$ is negative because the terms containing higher
powers of ${\frac{\delta{M}}{M_{c}}}$ are positive and they dominate  the other terms (which are negative,
since $\delta{M}$ is arbitrary). For large variations of $\delta{M}$ of any sign, we have $F < 0$, then we
deduce that $\sigma(M,t)$  is positive, and thus conclude that this PBH migrates to the evaporating regime.

Now, even if the variation $\delta{M}$ is assumed to be small and
negative, we could imagine that
$F(M_{c},\delta{M})$ may change its signal and become positive. We
shall prove that this is never the case. If the net variation in $M$ is small,
we can ignore the first term in the r.h.s. of eq.(40) and integrate to give

\be
\sigma(M,t_{i}+\Delta{t})\sim{+\biggl({\frac{c}{{t_{i}}}}\biggr)}^{3}\varrho_{ra
d}(t_{i})\delta{t} \,\, ; 
\ee

where $\delta{t}$ is a small time interval satisfying
$\delta{t} \ll \frac{27\pi{G^{2}}{c^{3}}}
g(M)\Delta{\varrho_{rad}}$ with
$g(M)={\biggl[{4A^{2}/{M}^{7}-A B \varrho_{rad}}\biggr]}^{-1}$.

In order to show that a PBH in an initial equilibrium $M = M_{c}$ deviates
from it is enough to show that the first
derivative $\dot{\sigma}$ is non-zero at $M = M_{c}$ and is
always positive for arbitrary variations
$\delta{M}$. We proceed by steps.

Since that we are interested in the algebraic sign of
$\sigma(M,t)$ only, we change variables to absorb the coefficient ${\frac{c^{6}}{32\pi{G}}}$
in eq.(40) and rewrite $\sigma$ as

\be
\sigma(M,t)={\frac{A}{M^{4}}}-B\varrho_{rad}(t) \,\, . 
\ee

To prove that $\dot{\sigma}(M = M_{c},t)>0$ it is enough to see that its
derivative is just
$\dot{\sigma}(M,t)=-{\frac{A}{M^{5}}}\dot{M}-B\dot{\varrho}_{rad}(t)$.
However, the first term is null at $M=M_{c}$
and the second term is positive because we assumed that the universe
expands, then $\dot{\sigma}(M,t)> 0$ beyond
the critical mass. Suppose that the variation in $M$ is such that we write
$M=M_{c} + \delta{M}$ and the "final" mass
is below the critical mass by a significant amount, say $M_0$.
Therefore, the energy flux conserves its original sign from the
initial equilibrium. Now, let us prove that
the sign of $\dot{\sigma(M,t)}$ does not change if the small variation
$\delta{M}$ is negative.
For this purpose, we write
$\dot{\sigma}(M,t)=-{\frac{A}{M^{5}}}\biggl({-{\frac{A}{M^{2}}}}+B\varrho_{rad}M^{2}
\biggr)-B\dot{\varrho}_{rad}(t)$.
If we substitute $M=M_{c}+ \delta{M}$ taking into account that this change is
very small, we can expand this expression and arrive at

\be
\dot{\sigma}(M,t)=-{\frac{A}{M^{5}}}\biggl[{\frac{A}{{M_{c}}^{3}}}+2M_{c}B\varrho_
{rad}(t)\biggr]\delta{M} > 0 \,\, .
\ee

Therefore, the rate of change of $\sigma(M,t)$ drives the hole to the evaporation. To see
that this condition of eq.(47) is enough we may calculate
${\frac{d\sigma}{dt}}(M_{c}+\delta{M})>0$, thus implying that
${d\sigma}(M_{c}+\delta{M})>0$ or
$\sigma(M_{c}+\delta{M})-\sigma(M_{c})>0$. This in turn means that
$\sigma(M_{c}+\delta{M})>0$, since $\sigma(M_{c})=0$ and $\sigma(M)>0$
happen in the evaporation region, or when the PBH losses much more energy than it is
able to absorb from the environment. In the opposite case,
if the mass variation is large but negative, we have
$\sigma(M=M_{c}+\delta{M})={\frac{A}{M^{4}}}-B\varrho_{rad}(t)\sim{\frac{A}{M^{4}}}>
0$, and the first term dominates because $\xi<<1$ in this case. This
completes the analysis for all negative variations $\delta{M}$ around the critical mass.

Now, we have in mind the general behaviour of the PBH mass evolution as the
universe expands. From the eq.(46) we deduce that

\be
\dot{\sigma}(M,t)=-{\frac{4A}{M^{5}}}\dot{M}-B\dot{\varrho}_{rad}(t) \,\, .
\ee

For subcritical black holes,  $\dot{M}<0$ and we know that
$\dot{\varrho}_{rad}(t)< 0$ due to
cosmological expansion. Then, for subcritical PBHs is immediate that
$\dot{\sigma} > 0$.

In the opposite case of supercritical black holes, the first term is
$-{\frac{4A}{M^{5}}}\dot{M} < 0$. Then, the
algebraic signal of $\dot{\sigma}(M,t)$ is determined by the difference
$-B\dot{\varrho}_{rad}(t)
-{\frac{4A}{M^{5}}}\dot{M}$. However, we found in section 1 that the
relative
mass gain is strictly limited to a few percent. This means that the
second term is negligible and the algebraic signal of $\dot{\sigma}(M,t)$
remains positive.

The algebraic sign of $\dot{\sigma}(M,t)$ is always positive in
"natural" circumstances. Some paths where $\dot{\sigma}(M,t) < 0$ are not
{\it a priori} excluded, but these paths correspond to very
special conditions where $\dot{\varrho}_{rad} > 0$ or the negative
variation in $\varrho_{rad}(t)$ is very small. For realistic
models of the radiative era, we know that
$\varrho_{rad}(t)\propto{t}^{-2}$, then, this variation is not
small compared to PBH mass gain. All those paths where
$\dot{\sigma}(M,t)< 0$ are entropically disfavoured.

In the general case when the universe expands, the natural
evolution for a PBH in the radiative era is to follow the type of
(almost) horizontal path drawn in the Figure 1, before crossing the critical mass 
curve. There is a small
mass gain along these paths, as we explained previously. The paths are
thermodynamically favourable and terminate at the instantaneous
equilibrium point $M_{c}(t = t_{c})$ at the crossing time
$t_{c}(M_{i})$ (marked with a cross). Now, since that $\dot{\sigma(M,t)}>0$ at the
critical mass, the deviation of equilibrium is unavoidable. The
PBHs follow the type downward path afterwards, where $\sigma(M,t)>0$
and its derivative are positive as already discussed. In other words,
the Hawking evaporation will eventually 
dominate the classical absorption.

In the general case we should analyze the full behaviour of

\be 
\sigma(M(t), t \gg t_{c})=\sigma(M_{i},t_{i})+{\frac{A{c}^{6}}{32{\pi}{G}^{2}}}
\biggl[{{\frac{1}{{M}^{4}(t)}}-{\frac{1}{{M_{i}}^{4}}}}\biggr] 
-{\frac{27{c}^{3}}{32}}\Delta{\varrho_{rad}(t)} \,\, ; 
\ee

\bigskip

where $M(t)$ is a solution of
$\dot{M}=-{\frac{A}{M^{2}}}+B \varrho_{rad}(t){M}^{2}$. If the initial 
conditions are 
given by $M_{i} > M_{c}(t_{i})$, $\varrho_{rad}(t)\propto{t}^{-2}$ (and it is
implicit that
$\varrho_{pbh}(t)<<\varrho_{rad}(t)$ as before); 
and substituting the appropriate solution for $M(t)$ into eq.(49) above, 
we deduce afte taking
$t>>t_{c}(M_{i})$ that  $\sigma(M(t>>t_{c}),t)>0$.
Therefore such a black hole crossed the critical mass curve and is 
now below the critical mass at $t$. The hole is then hotter than the environment 
and must evaporate.

All these thermodynamical considerations show quite generally that an 
explosive growth of PBHs in the radiation era is forbidden. This conclusion 
in unavoidable when the problem is analyzed using carefully the critical 
mass. The validity of Sivaram's solution would also lead to the conclusion that 
an initially sub-horizon PBH would have to cross the horizon in a short time given by 
$\Delta t \simeq \frac{(M_{hor} - M_{pbh})}{3 \times 10^{38} g} \, s$. This 
kind of behaviour was already ruled out by a set of calculations \cite{ste} using full General 
Relativity. Independently of the present results and former work, we are able to prove that this strong growth is impossible if Bekenstein's limit \cite{bek} holds (P.S. Cust\'odio and J.E. Horvath, in preparation).

\section{Conclusions}

We have analyzed in this work some features of the evolution of
PBH mass in the radiation-dominated era. We found that within FRW
models with null curvature there is no room for an explosive
growth of their masses, which would need relativistic speeds of
the gravitational radius. This result is in line with the previous
analysis of Refs.\cite{ste} and \cite{NOS}. Indeed the very modest value of the
maximum fractional gain ($0.04$) argues against the possibility
discussed in Ref.\cite{Siv}. This means that essentially all the methods
applied to limit their abundance make sense since evidence for
these objects could lead to very valuable insights. We have also
addressed in some detail two ways to preclude evaporation: 
a large proper motion
altering the absorption-emission properties of the semiclassical
PBHs and the possibility that they could be "dragged" and stay on
the critical mass curve. In the first case we have given an
ultrarrelativistic motion of the holes at their very birth
($\gamma \geq 36$) ; and demonstrated that the second situation is
one of unstable equilibrium, and therefore that PBHs do cross the
critical mass curve as discussed in \cite{NOS} to evaporate (Fig.1).
Other forms to avoid the onset of the evaporation might in
principle be possible, but are not known to us.

\section{Acknowledgements}

Both authors wish to thank the S\~ao Paulo State Agency FAPESP for
financial support through grants and fellowships. J.E.H. has been
partially supported by CNPq (Brazil).

\clearpage

\section{Appendix: a general argument against a linear growth}

We shall now discuss further consequences of a linear growth of the 
mass as derived in Ref. \cite{Siv}. If we assume that it is correct and 
substitute $M(t) = M_{i} (t/t_{i}$ back into 
$\dot{M} (t) = B \varrho_{rad}(t) M^{2}(t)$ and impose that 
$\varrho_{rad}(t) = \varrho_{rad} (t_{i}) (t_{i}/t)^{2}$, we 
obtain for the r.h.s.

\be
{\frac{dM}{dt}} = B \varrho_{rad}(t_{i}) = constant \,\, ;
\ee

for arbitrary  times. Using again the same expression $M(t) = M_{i} (t/t_{i})$ and 
rearranging yields

\be
\frac{M_{i}}{t_{i}} = B M_{i}^{2} \varrho_{rad}(t_{i}) \,\, ; 
\ee

which shows that a linear solution would require a definite value of the 
radiation density at the formation time. Since initial black hole masses 
and the environment radiation density must be independent, we conclude that 
a linear growth is unphysical.

We can generalize the above fine tuning constraint for arbitrary 
conditions provided reasonable physics still holds. 
Suppose that we
do not know the behaviour of the radiation density at all. Then,
it would be reasonable to guess for it some function of the form
$\varrho_{rad}(t)=\varrho_{rad}(t_{i})F(t/t_{i})$.

If we substitute this guess into eq.(17), we obtain

\be
(1-1/u)=M_{i}B\varrho_{rad}(t_{i})\int_{1}^{u}du^{\prime}F(u^{\prime}) \,\, ;
\ee

where $u=(t/t_{i})$. A power-law ansatz $F(u)={u}^{m}$ for some $m$ gives

\be
(1-1/u)={\frac{1}{(m+1)}}[BM_{i}\varrho_{rad}(t_{i})](u^{(m+1)}-1) \,\, .
\ee

The solution for $m=-2$ displays the same unphysical fine tuning
as before, since that it requires $BM_{i}\varrho_{rad}(t_{i})=1$.

On general grounds there is no physical solution for 
$m$ different from $-2$ which does not require some fine tuning between the initial
mass $M_{i}$ and the radiation density at this moment. Therefore,
we conclude that the solution $M(t)\propto{t}$ is ruled out.

\clearpage

\section{Figure captions}
\bigskip
\noindent

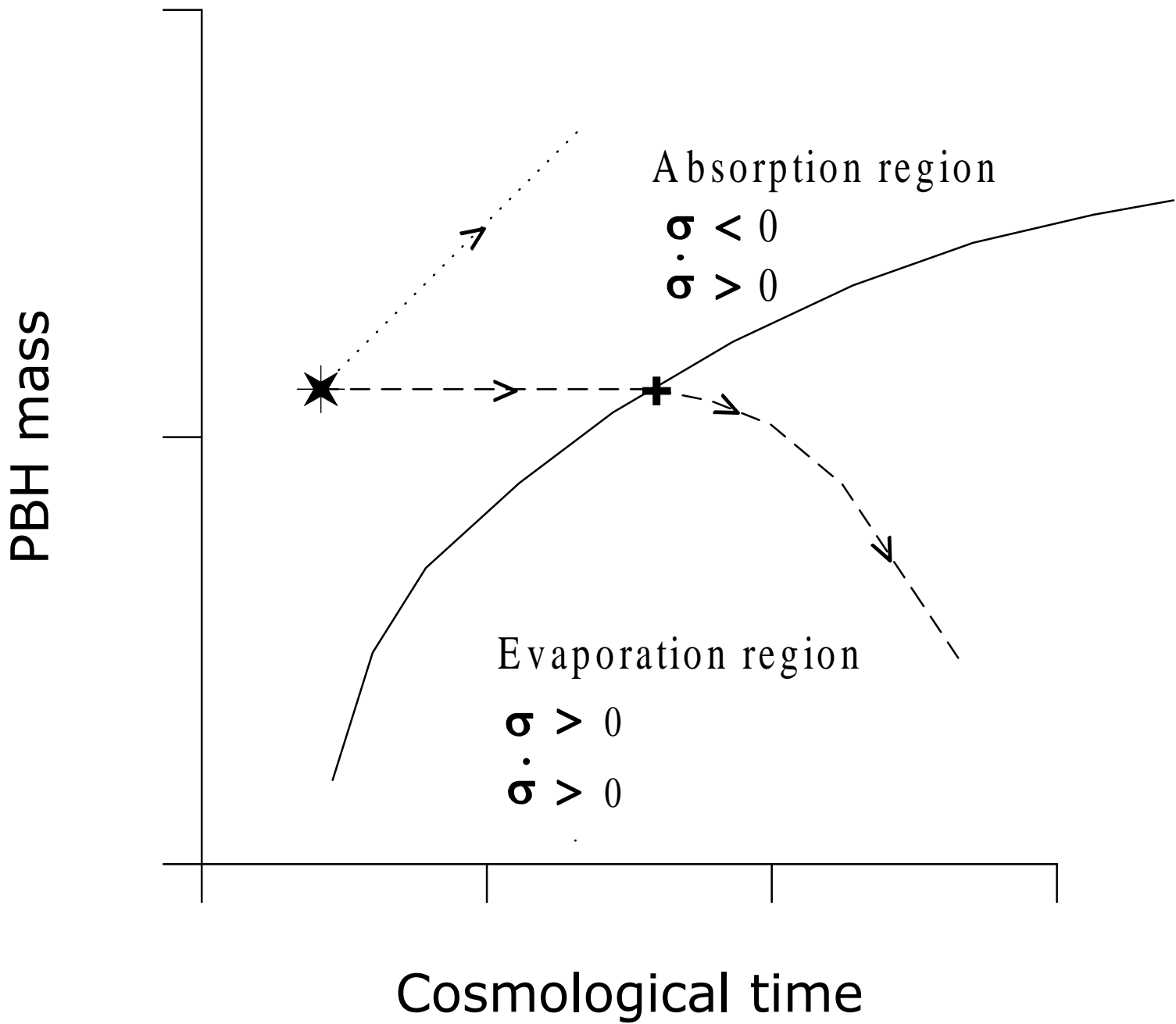
Figure 1. Evolution of PBHs in the absorption and evaporation regimes

The type of paths of a generic PBH are sketched in the PBH mass vs. cosmological time 
plane. The PBH is formed at the point marked with the star and undergoes a very small 
mass gain as long as its mass is larger than the critical mass (solid line), thus 
following an almost perfect horizontal path (long dashed) until the crossing time at the point 
marked with the cross. Since, as discussed in Section 3, the equilibrium is instantaneous, 
the PBH must drift to the evaporation region where the Hawking radiation will 
cause a decrease of the PBH mass and thus an evolution along the downwards bending path as 
indicated. Note that the mass as a function of time has a maximum on 
the critical mass curve; given that $\dot{\sigma}$ has the inverse sign than $\ddot{M}$. 
The path claimed in Ref. 3 is depicted with a dotted line and 
would need unlikely conditions as discussed in the text.

\end{document}